\documentclass[aps,prl,reprint,floatfix,showpacs]{revtex4-1}
\usepackage{amssymb,amsmath}
\usepackage{graphicx}

\newcommand{\psLabelSize}{1.}
\newcommand{\psTickSize}{.5}
\newcommand{\psAxesLabelSize}{.7}

\usepackage{color}

\newcommand{\abs}[1]{\left| #1 \right|}
\newcommand{\vv}[1]{\mathbf{ #1 }}

\newcommand{\nPartTypes}{K}
\newcommand{\V}[2]{V^{#1#2}}

\newcommand{\figref}[2]{Fig.\ \ref{#1}#2}
\newcommand{\figsref}[2]{Figs.\ \ref{#1} and \ref{#2}}

\usepackage{psfrag}

\begin{document}

\title{Predicting self-assembled patterns on spheres with multi-component coatings}

\author{E. Edlund}
\author{O. Lindgren}
\author{M. \surname{Nilsson Jacobi}}%
\email{mjacobi@chalmers.se} 
\affiliation{Complex Systems Group, Department of  Energy and Environment, Chalmers University of Technology, SE-41296 G\"oteborg, Sweden}
\date{\today}%

\begin{abstract}
	Interactions between the components in many-body systems can give rise to spontaneous formation of complex structures.
	Usually very little is known about the connection between the interactions and the resulting structure. 
	Here we present a theory for self-assembling pattern formation in multi-component systems, formulated as an analytic technique that predicts morphologies directly from the interactions in an effective model.
	As a demonstration we apply the method to a model of alkanethiols on spherical gold particles, successfully predicting its morphologies and transitions as a function of the interaction parameters. 
	This system is interesting because it has been suggested to provide an effective route to produce patchy colloids. 
\end{abstract}


\maketitle

Spontaneous pattern formation is a fascinating and common phenomenon in nature. The underlying mechanisms are typically nonlinear and therefore difficult to analyze analytically.
Recently our understanding of pattern formation has increased significantly by applying group theory to classify different patterns and relate their appearance to the symmetries of the governing equations.
Quantitative predictions of the transition between different patterns is however still typically outside the reach of analytic techniques, and can for example not be addressed with linear stability analysis.
This is especially true in many-body systems with many different types of particles and interactions where relatively complex morphologies can emerge.

An example of such a system is nanoparticles coated with mixtures of ligands that phase separate; these display various morphologies depending on the properties of ligands and nanoparticles~\cite{jackson_spontaneous_2004}.
Here spontaneous pattern formation offers an interesting route to fabrication of patchy colloids, particles with patches on their surfaces resulting in anisotropic interactions. 
Patchy colloids have received much interest as model systems for studying rheology, gel-arrest, and self-assembly in protein and polymer systems~\cite{bianchi_patchy_2011,yi_progress_2013}.
They can also be used as potential building blocks for directed self-assembly of colloidal particles themselves in pursuit of new meta materials~\cite{Glotzer:2004ce,pawar_fabrication_2010,Doppelbauer:2012kf}.
Motivated by this a number of recent numerical studies have started to map out the morphologies expected to self-assemble in binary~\cite{singh_entropy-mediated_2007}, ternary~\cite{pons-siepermann_design_2012a}, and quaternary~\cite{pons-siepermann_design_2012b} mixtures on spheres. 
However, both prediction and characterization of the morphologies appearing in these systems are difficult. The nature of the patterns formed in experiments has for example been subject to an extensive debate~\cite{jackson_spontaneous_2004,cesbron_stripy_2012,yu_response_2012}.

In this Letter we present theory for predicting pattern formation through self-assembly in multi-component systems, \emph{via} effective interaction potentials.
From the interactions we can determine the features that define the patterns, such as the type of morphology and its characteristic wavelength(s).
This also allows us to analytically predict transition points, defined as abrupt changes in the morphologies caused by variations of the interaction parameters. 
To verify the theory we apply it to a model of alkanethiol-on-gold systems with effective isotropic interactions and compare predicted morphologies with states obtained from Monte Carlo simulations. 


Alkanethiol-on-gold patchy colloids are examples of a class of systems where metal nanoparticles are covered by self-assembled monolayers formed by mixtures of thiol surfactants. 
The alkanethiols bind semicovalently to the surface of a gold nanoparticle \emph{via} their thiol head groups and are chosen to have different end groups, causing them to be immiscible and thus to prefer to segregate by type. 
However, attractive entropic forces arise from an increase in available free volume for ligands with shorter or less bulky neighbors~\cite{singh_entropy-mediated_2007}.
The competition between these effects gives a degree of mixing, in two-component systems resulting in the formation of stripes, depending on the length of the chains and strength of the interactions~\cite{jackson_spontaneous_2004}.
If more than two types of thiol surfactants are mixed, a wealth of different morphologies can appear~\cite{pons-siepermann_design_2012a,pons-siepermann_design_2012b}.

We study a minimalistic model of the same system using isotropic point particles to represent the alkanethiols. 
It can be viewed as a simplified version of some effective potential obtained by integrating over the internal degrees of freedom of the molecules~\cite{likos_effective_2001}.
The following briefly describes the model, which we here take as given. 
In a forthcoming paper~\footnote{E. Edlund, O. Lindgren, and M. Nilsson Jacobi, in preparation (2013)}, we will give a detailed motivation and show that when we perform Monte Carlo annealing for a wide range of parameters the model turns out to capture most morphologies and transitions of the original system.
It will thus serve as a useful test case for our theory.


We consider $\nPartTypes$ particle types, $\alpha \in \{1,\dots, \nPartTypes\}$, to which we assign length parameters $L_\alpha$ (to be viewed as abstractions of the lengths of the alkanethiols).
We posit particles of the same type to simply interact \emph{via} a hard-core potential with a diameter $\sigma_0$,
\begin{equation}
	\tag{\addtocounter{equation}{1}\arabic{equation}a}
	\label{alkanethiol-model_a}
	\V{\alpha}{\alpha}(r)= \begin{cases} 
		\infty, & \mbox{if } r < \sigma_0 \\
						0, & \mbox{otherwise,}
	\end{cases}
\end{equation}
while the potential between particles of different types is taken to be
\begin{equation}
	\tag{\arabic{equation}b}
	\label{alkanethiol-model_b}
	\V{\alpha}{\beta}(r)= \begin{cases} 
		\infty, & \mbox{if } r < \sigma_0 \\ 
						1, & \mbox{if } \sigma_0 < r <  \sigma_1 \\ 
-\epsilon, & \mbox{if } \sigma_1 < r <  \abs{L_\alpha - L_\beta} \\ 
							 0, & \mbox{otherwise.}
	\end{cases}
\end{equation}
This includes a short-ranged soft shoulder potential to achieve immiscibility and a  square-well potential representing the entropic attraction from the alkanethiol length mismatch.


We now formulate a theory that allows us to study the pattern formation in systems  described by an effective Hamiltonian with isotropic interactions (such as the alkanethiol model just described). In this method the original system is approximated with a model with relaxed constraints. The approximate model can be solved analytically and the solutions then used in combination with the original constraints to predict the pattern formation in the full system.

Consider a generic Hamiltonian with $\nPartTypes$ different particle types, interacting with different isotropic pairwise interactions. 
We describe this with a Potts-like model~\cite{baxter_exactly_1982} on a lattice, where each lattice site $i \in \{1, \dots, N\}$ is occupied by a particle of type $\alpha \in \{1, ... , \nPartTypes\}$%
~\footnote{
Here the careful reader notes that it is impossible to construct a uniform mesh on a sphere (except for the meshes of 4, 6, 8, 12, or 20 points corresponding to the Platonic solids).
However, our theory does not need a uniform mesh.
For the numerical spectra calculations, we use a configuration obtained by Monte Carlo annealing~\cite{Note4} for $N=880$ particles on a sphere with radius $9$.
Using different configurations makes negligible difference for the spectra.
}%
. 
The state can then be described by a $N\times K$ matrix $\Pi_{i\alpha}$, which in our setting also can be viewed as a vector with two indices. The elements are 0 or 1 depending on whether a particle of type $\alpha$ is present at site $i$. 
We express the Hamiltonian in terms of $\Pi _{i \alpha}$ as
\begin{equation}
	\label{hamiltonian}
	H = \sum_{\alpha \beta}^\nPartTypes \sum_{ij}^N \Pi_{i\alpha} \, \V{\alpha}{\beta}_{ij} \, \Pi_{j\beta},
\end{equation}
where the interactions between a particle of type $\alpha$ at site $i$ with one of type $\beta$ at site $j$ is described by  the potential $\V{\alpha}{\beta}_{ij} =  \V{\alpha}{\beta}(\abs{\vec{r}_i - \vec{r}_j})$. 

To allow analytic treatment we approximate the discrete model with a generalization of the spherical model  (\emph{cf.}\ the procedure in~\cite{edlund_universality_2010}), where the discrete elements of $\Pi _{i \alpha}$ are replaced by continuous variables combined with a global constraint $\sum_{i\alpha} \abs{\Pi_{i\alpha}}^2 = N$.
In this relaxed formulation the ground states can be found analytically through a direct diagonalization of the interaction matrix.

\begin{figure}[t]
	\includegraphics[width= 0.35\textwidth]{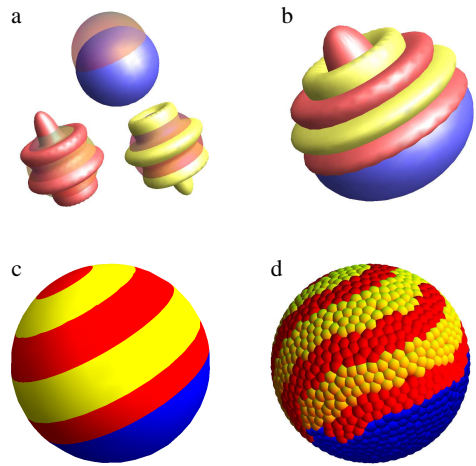}
	\caption{
		Obtaining patchy colloids from spherical harmonics.
		(a) Spherical harmonics with $l=1$ (blue) and $l=5$ (red, yellow); shown are real parts as radial perturbations from a sphere. (b) States in the continuous model are given by combinations of such harmonics determined by the energy spectrum (\figsref{multifig}{lengthfig}). Thresholding of these states (c) corresponds closely to low energy states obtained by Monte Carlo simulations (d).
	\label{eigenvectorIllus}}
\end{figure}

The diagonalization involves two steps. 
First we note that all interaction matrices ($V^{\alpha \beta}_{ij}$ with fixed $\alpha$ and $\beta$) describing systems with isotropic interactions are simultaneously diagonalized by eigenfunctions of the Laplace operator~\cite{edlund_universality_2010,nussinov_commensurate_2001}. 
On surfaces of spheres, these are spherical harmonics $Y^m_l(\theta,\phi)$ (\figref{eigenvectorIllus}{a}). 
By rotational symmetry, the energy is independent of $m$ and we thus drop this index when possible.

\begin{figure}[tb]
\includegraphics[width= 0.45\textwidth]{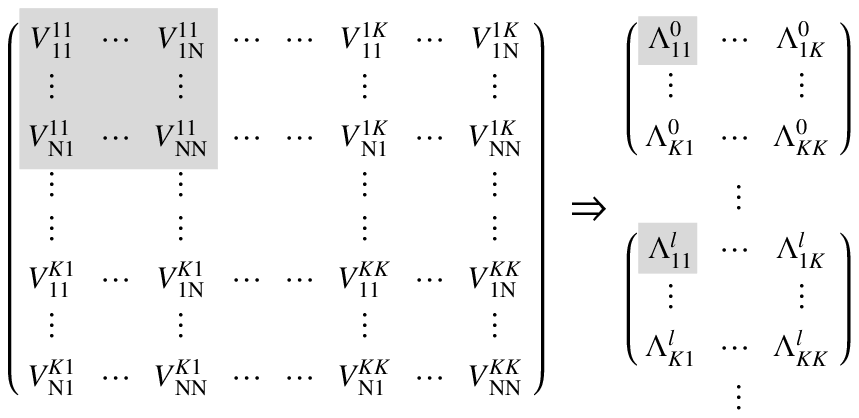}
\caption{\label{matrixIllustration}
The interactions can be written as a matrix with $\nPartTypes\times \nPartTypes$ blocks. We independently diagonalize each block and construct the smaller $\nPartTypes\times \nPartTypes$ matrices $\Lambda^l$ from the resulting eigenvalues.
$\nPartTypes$ is the number of particle types and $N$ the number of particles.
}
\end{figure}


\begin{figure*}[tb]
	\psfrag{a}[c][c][\psLabelSize][0]{(a)}
	\psfrag{b}[c][c][\psLabelSize][0]{(b)}
	\psfrag{c}[c][c][\psLabelSize][0]{(c)}
	\psfrag{E}[c][c][\psAxesLabelSize][0]{$E$}
	\psfrag{l}[c][c][\psAxesLabelSize][0]{$l$}
	\psfrag{1}[c][c][\psTickSize][0]{$1$}
	\psfrag{5}[c][c][\psTickSize][0]{$5$}
	\psfrag{9}[c][c][\psTickSize][0]{$9$}
	\psfrag{t}[cr][c][\psTickSize][0]{$-10$}
	\psfrag{T}[cr][c][\psTickSize][0]{$10$}
	\psfrag{s}[cr][c][\psTickSize][0]{$-30$}
	\psfrag{S}[cr][c][\psTickSize][0]{$30$}
	\includegraphics[width= 0.31\textwidth]{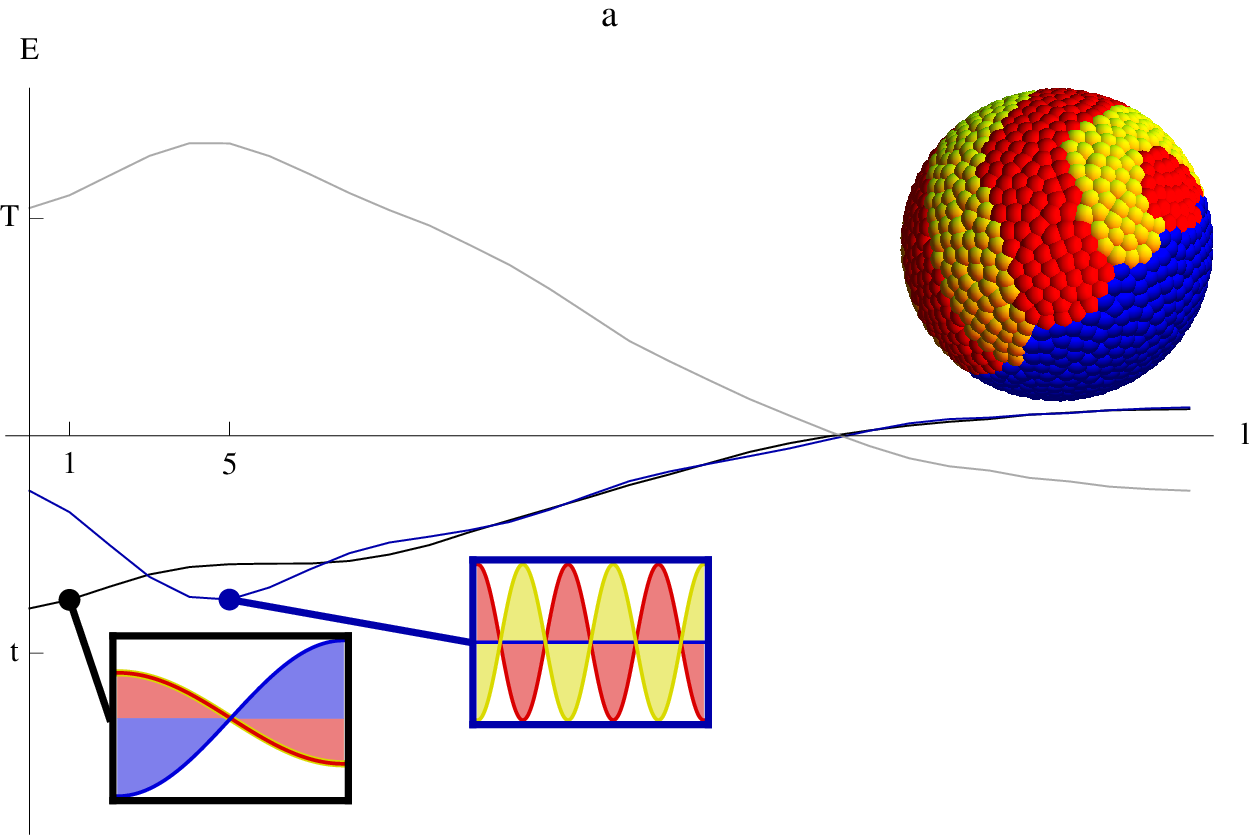}\hspace{5mm}
	\includegraphics[width= 0.31\textwidth]{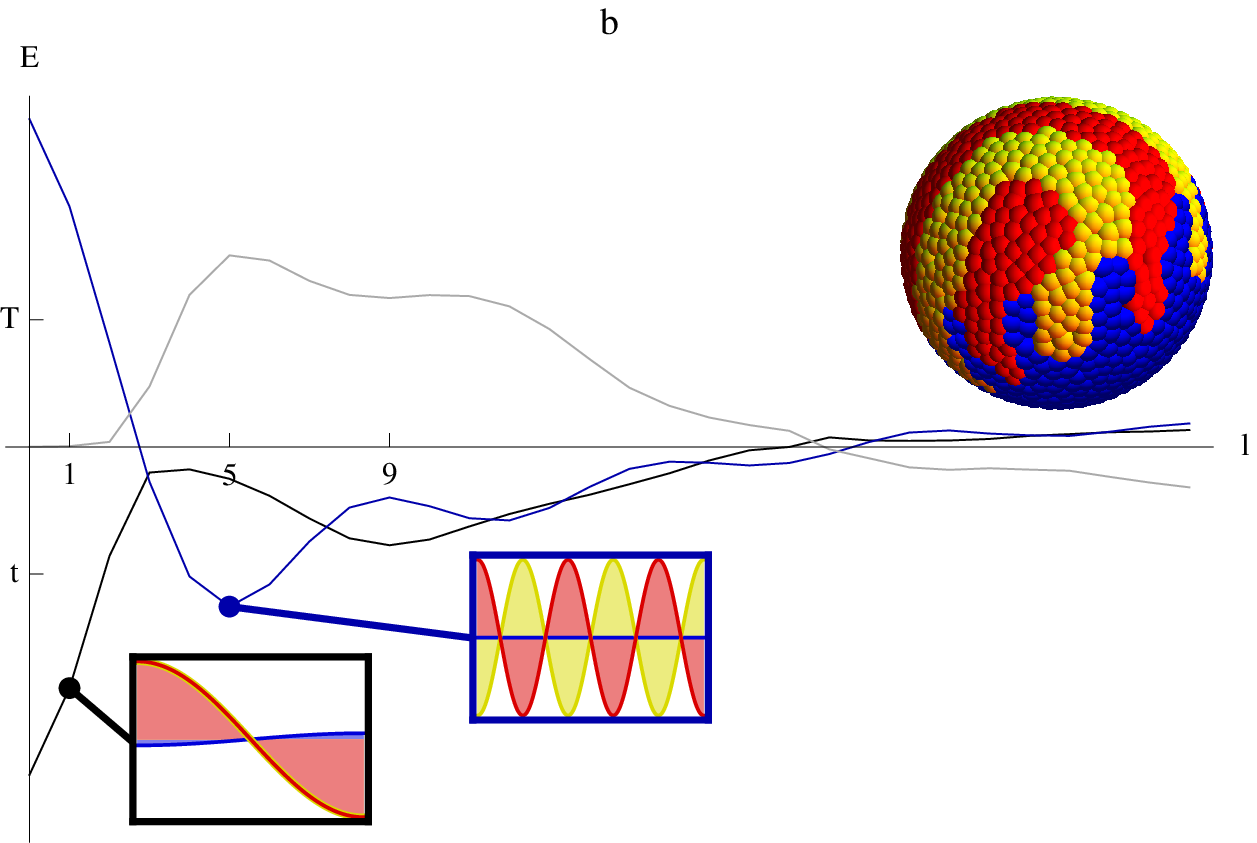}\hspace{5mm}
	\includegraphics[width= 0.31\textwidth]{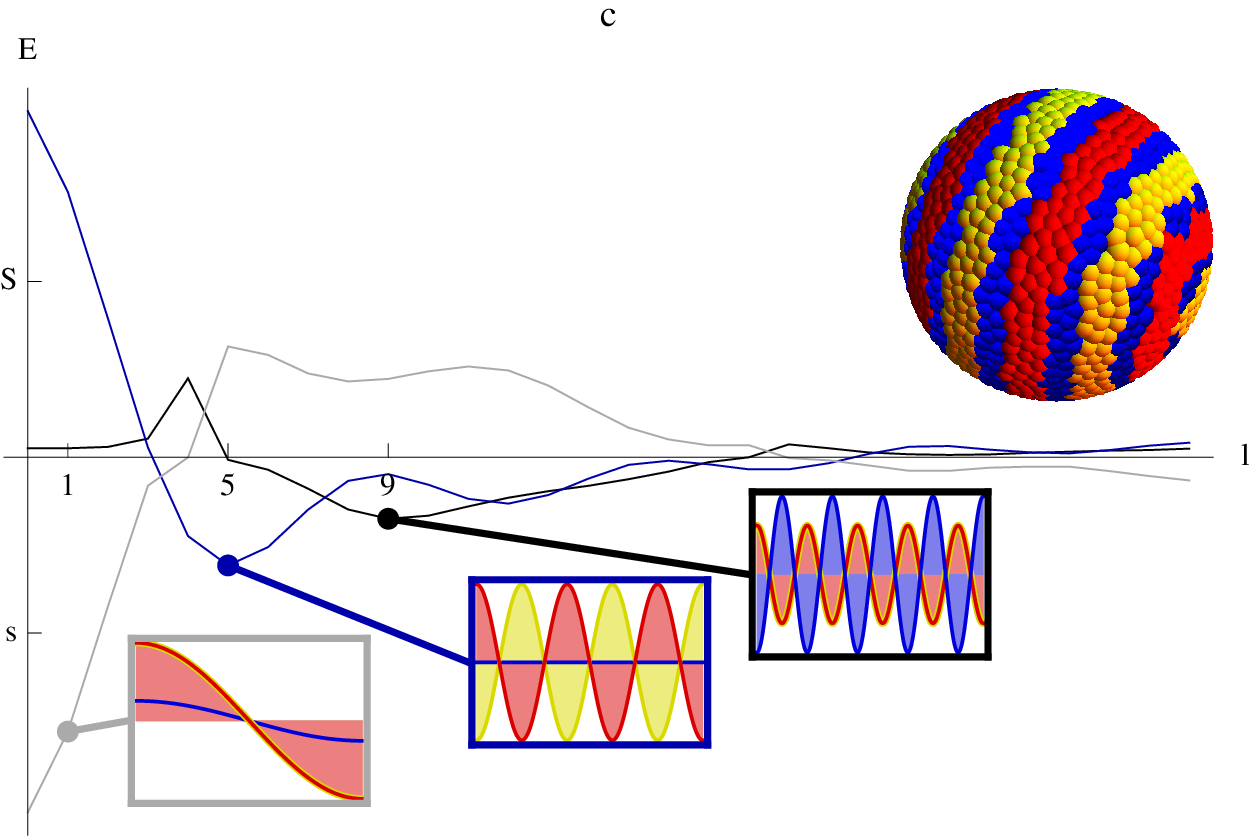}
	\caption{ 
		Energy spectra of the alkanethiol model with $\sigma_1=\sqrt{3}\sigma_0$, $L_{red} = 0$, $L_{blue} = 4$, and  $L_{yellow} = 8$ for  $\epsilon = 0.03$ (a), $\epsilon = 0.2$ (b), and $\epsilon = 0.4$ (c). 
		The insets show (top) results of Monte Carlo annealing~\cite{Note4} with stoichiometry 1:1:1 and (bottom) one-dimensional representations of states (Fourier modes with amplitudes and phases given by $\vv{a}_{l,i}$ and frequencies corresponding to $l$) for points discussed in the main text.
		The transition from striped Janus (a), \emph{via} a partially mixed state (b), to alternating stripes (c) occurs due to the sign change of the blue phase in the global minimum.
	\label{multifig}}
\end{figure*}

Let $\Lambda^{l}_{\alpha \beta}$ denote the eigenvalue of $\V{\alpha}{\beta}$ corresponding to the spherical harmonics $Y^m_l(\theta,\phi)$ and for each $l$ form the (symmetric) $\nPartTypes\times \nPartTypes$-matrix $ \Lambda^{l}_{\alpha \beta}$ (see \figref{matrixIllustration}{}).
We can then write the original Hamiltonian \eqref{hamiltonian} as
\begin{equation}
	H = \sum_{l,m} \vv{v}_{l,m}^T \Lambda^{l} \vv{v}_{l,m},
\end{equation}
where the original variables $\Pi_{i\alpha}$ are now rotated into a set of $\nPartTypes$-dimensional vectors $\vv{v}_{l,m}$.
The last step consists of diagonalizing these $\Lambda$-matrices independently.
This gives $\nPartTypes$ eigenvalues for each $l$.
The collection of all these eigenvalues forms the energy spectrum $\{E_{l,k}\}_{l,k}$ (see \figsref{multifig}{lengthfig}), a spectrum with $\nPartTypes$ branches ($k\in\{1,\dots,\nPartTypes\}$)%
~\footnote{The reader may note the similarity between our diagonalization procedure and the standard approach to calculating lattice phonon spectra~\cite{ashcroft_solid_1976}, though the interpretation of the results is different.}.
The ground state energy $E_{l_{min},k_{min}}$ of the Hamiltonian \eqref{hamiltonian} can now be found as the minimum of this spectrum, at some $l_{min}$ and some branch $k_{min}$.

The diagonalization of each $\Lambda$-matrix also gives $\nPartTypes$ eigenvectors $\vv{a}_{l,k}$.
To construct the ground state of the approximate model we use the eigenvector corresponding to the lowest eigenvalue in the spectrum as the amplitude of the spherical harmonic, constructing the state $\Pi = \left[ \left(\vv{a}_{l_{min},k_{min}}\right)_1 \cdot Y_{l_{min}},\dots,  \left(\vv{a}_{l_{min},k_{min}}\right)_\nPartTypes \cdot Y_{l_{min}} \right]$, \emph{i.e.}, each column of the state matrix $\Pi$ is a spherical harmonic with orbital number $l_{min}$ and amplitude given by the corresponding element of $\vv{a}_{l_{min},k_{min}}$ (see \figref{eigenvectorIllus}{b} for an example with two $l$:s). 

\begin{figure*}[tb]
	\psfrag{a}[c][c][\psLabelSize][0]{(a)}
	\psfrag{b}[c][c][\psLabelSize][0]{(b)}
	\psfrag{c}[c][c][\psLabelSize][0]{(c)}
	\psfrag{E}[c][c][\psAxesLabelSize][0]{$E$}
	\psfrag{l}[c][c][\psAxesLabelSize][0]{$l$}
	\psfrag{1}[c][c][\psTickSize][0]{$1$}
	\psfrag{5}[c][c][\psTickSize][0]{$5$}
	\psfrag{7}[c][c][\psTickSize][0]{$7$}
	\psfrag{t}[cr][c][\psTickSize][0]{$-10$}
	\psfrag{T}[cr][c][\psTickSize][0]{$10$}
	\includegraphics[width= 0.31\textwidth]{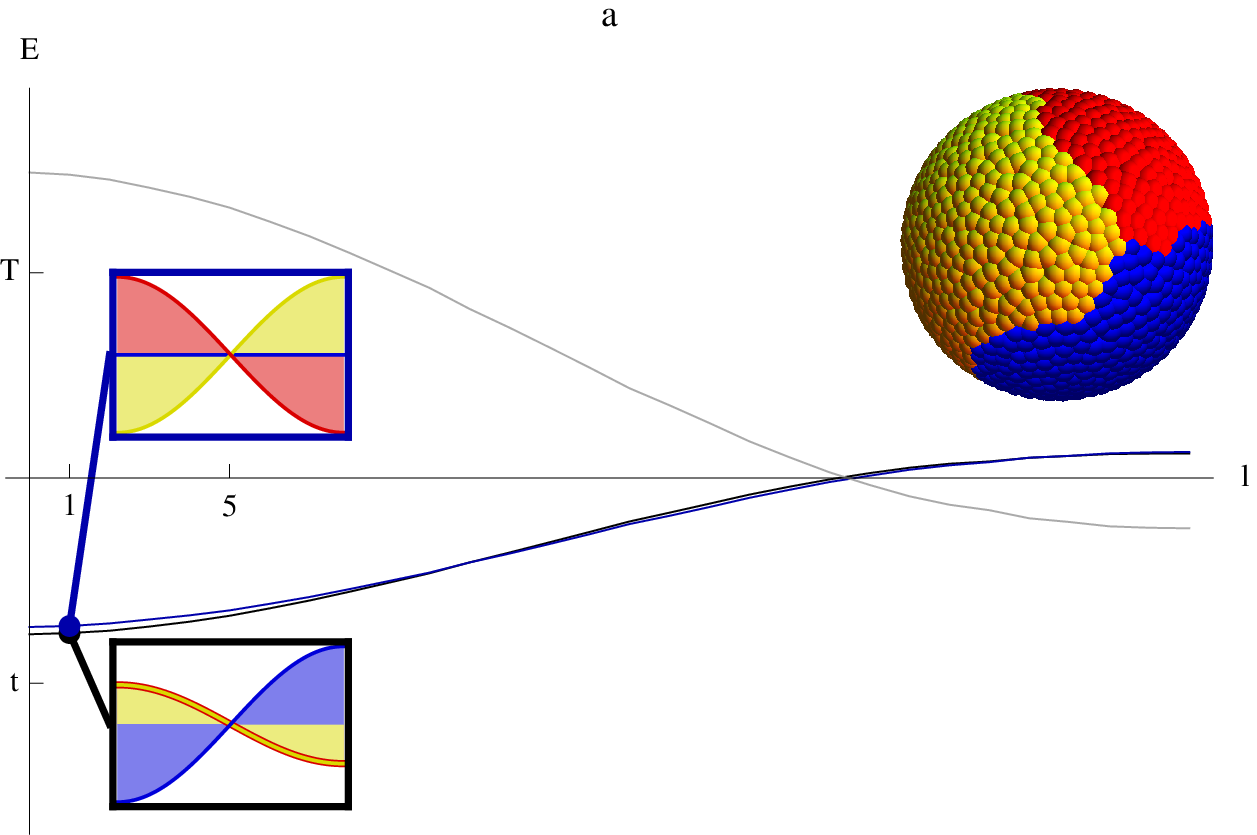}\hspace{5mm}
	\includegraphics[width= 0.31\textwidth]{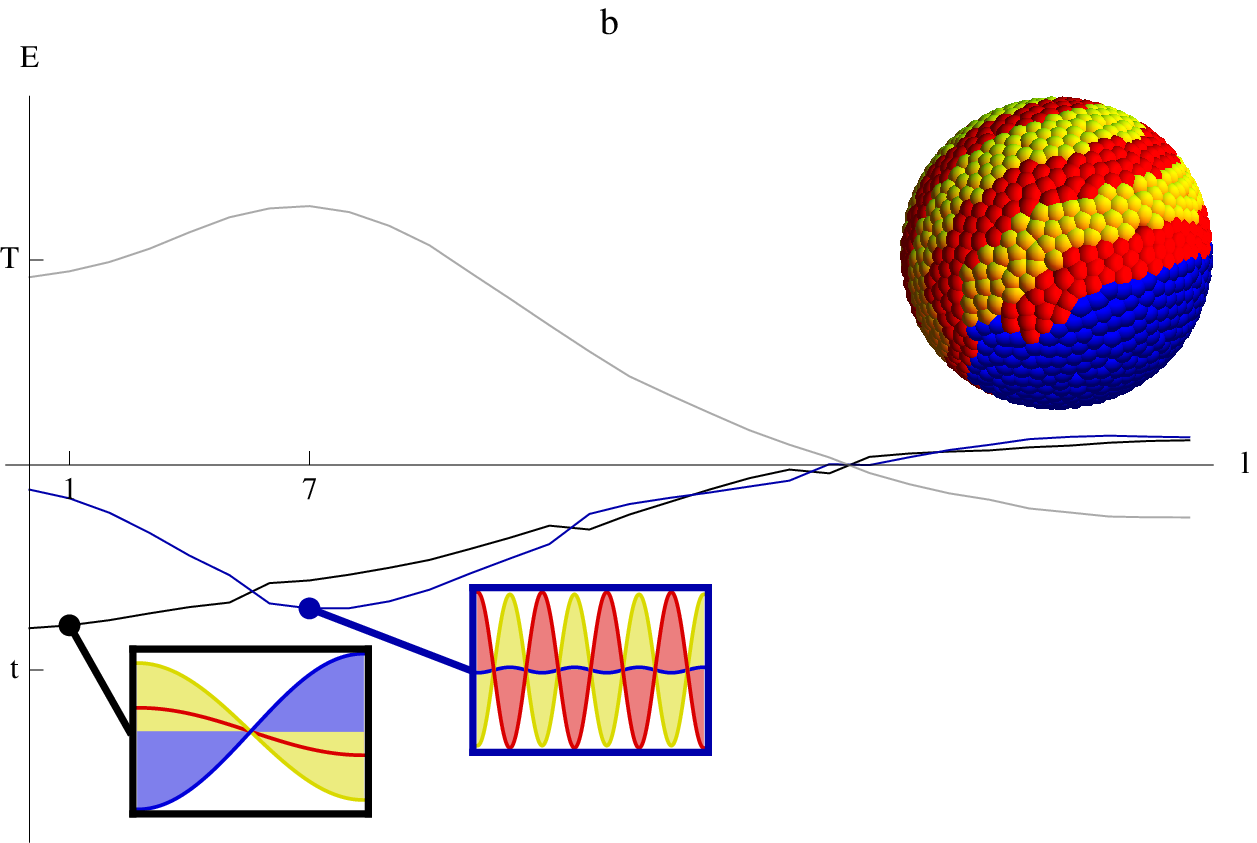}\hspace{5mm}
	\includegraphics[width= 0.31\textwidth]{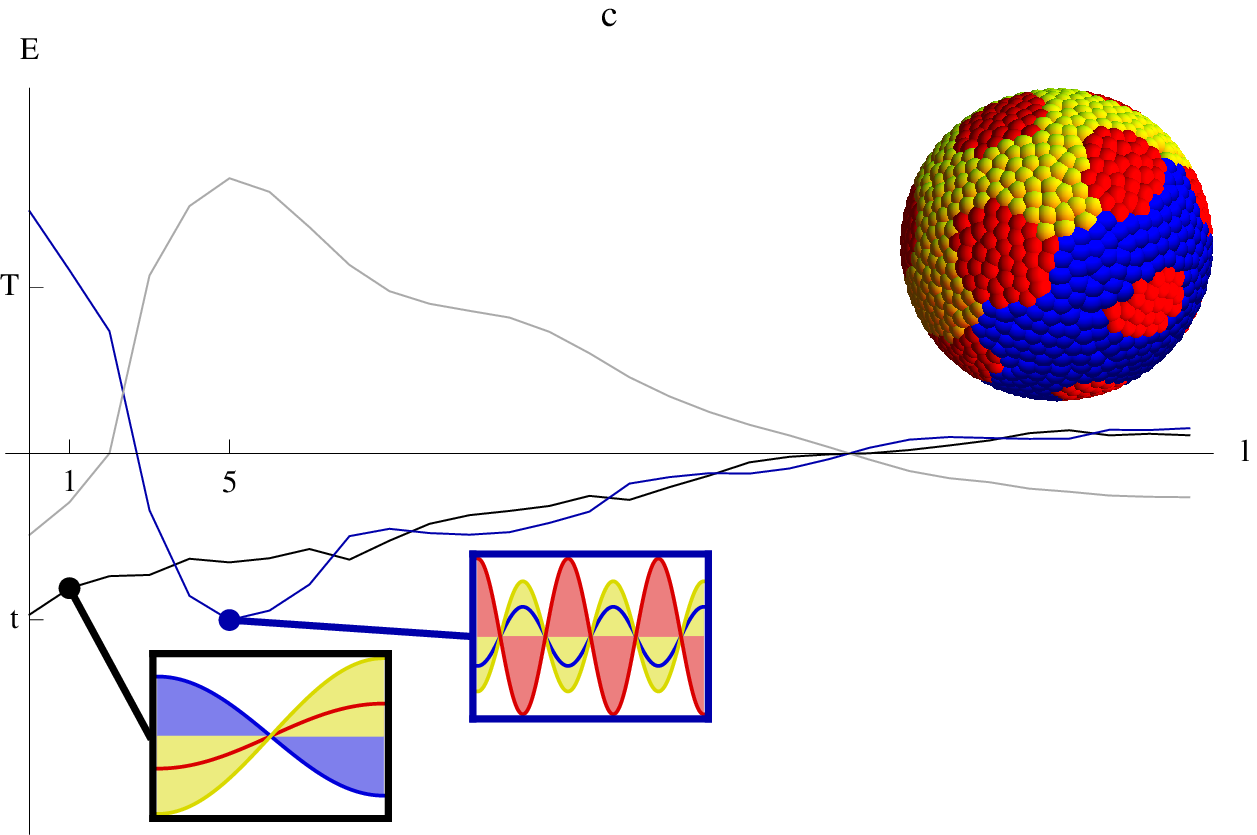}
	\caption{ 
		Energy spectra of the alkanethiol model with $\sigma_1=\sqrt{3}\sigma_0$, $\epsilon = 0.1$, $L_{red} = 0$, and $L_{blue} =L_{yellow} -1$ with $L_{yellow}=2$ (a), $L_{yellow}=5$ (b), and $L_{yellow}=8$ (c).
		Breaking the degeneracy that causes the Cerberus particle (a) changes it into a striped Janus (b). A cross-over of the global minimum changes the morphology to a spotted Janus (c).
	\label{lengthfig}}
\end{figure*}

This construction gives analytical solutions to the continuous model.
To produce predictions for particle systems these solutions must be re-interpreted in the context of the original model. 
With two particle types this interpretation is straightforward, involving only a rounding of the continuous variables of a single state to discrete values~\cite{edlund_universality_2010}.
In the cases we are now considering, with several particle types, the mapping is done on combinations on states and choosing these requires more detailed attention. 

First, for an eigenvector to be physical it must describe a separation of the particle types into two groups with opposing phase, \emph{i.e.}, the amplitudes $\vv{a}_{l,k}$ cannot all have the same sign since such solutions cannot be instantiated in the particle model.
Thus, to predict particle states we must exclude these unphysical solutions (in \figsref{multifig}{lengthfig} we draw the corresponding branches in light gray).

Second, if we consider a particle model where the stoichiometry is fixed, this must be taken into account. 
For example, the global minimum of the spectra might prescribe a state with alternating stripes of red and yellow particles (as in \figref{multifig}{c}), but if we want to predict the ground state of a system with three particle types in equal proportions we must also include the lowest energy mode describing how the blue particles separates from the rest. 
In general each physical branch represent a separation of the constituents into two groups. 
The complete pattern for a given stoichiometry is defined by a sufficient number of minima to describe how each particle type is separated from the others.

In summary, to predict the low-energy states of a multi-component particle model, we use the following procedure:
\begin{enumerate}
\item Construct the interaction potentials $\V{\alpha}{\beta}(r)$ between each pair of particle types $\alpha$ and $\beta$.
\item Take the spherical harmonics analogues of Fourier transforms of the interaction potentials to obtain the pair-wise energy spectra $\Lambda_{\alpha \beta}^{l}$ (\figref{matrixIllustration}{}).
\item For each $l$, diagonalize $\Lambda^{l}$, giving the full spectrum with its $\nPartTypes$ branches together with their amplitudes and phases $\vv{a}_{l,k}$ (\figsref{multifig}{lengthfig}).
\item Identify the minima of the spectrum's branches, excluding unphysical ones, and include enough of them to express the postulated stoichiometry.
\item Combine the spherical harmonics corresponding to these minima and map them to a prediction for the particle system (\figref{eigenvectorIllus}{b-c}).
\end{enumerate}


We will now demonstrate the theory on the alkanethiol model described above, in the special case of three particle types with equal stoichiometry.
We will compare the theory's predictions with Monte Carlo simulations of the model%
~\footnote{
All particle configurations displayed in this Letter are generated using Monte Carlo annealing of $N=1500$ particles restricted to the surface of a sphere with radios $R=12\sigma_0$. 
Trial moves consists of either, with equal probability, interchanging the position of two particles or choosing a new random position on the surface of the sphere for a single particle.
The annealing is done for $10^4$ sweeps at each of the temperatures $T\in \{1000, 100, 10, 3, 1, 0.3, 0.1\}$, where a sweep is $N$ trials.
}%
, and show how low-energy states can be predicted as well as transitions between them.

Consider first \figref{multifig}{a}.
It shows the energy spectrum for the alkanethiol model for a certain set of parameters.
The global energy minimum is attained by the black branch at $l=0$ which describes a non-mixed state. 
However, this is not allowed by the stoichiometry and instead the mode at $l=1$ has to be used, describing a large blue domain being in opposing phase to a red-yellow one.
To find out what happens within the red-yellow domain, we then look for the minimizing point where eigenvector ${\vv{a}_{l,i}}$ corresponds to red and yellow being in opposing phases.
This is found at $l=5$, prescribing red and yellow stripes with a wavelength allowing for approximately three stripes of each type.
The prediction is thus of a Janus particle decorated with stripes on one side.
Performing Monte Carlo annealing shows that the low energy states of the particle model indeed has a blue domain and a red-yellow striped one with the indicated wavelength, as shown in the inset of \figref{multifig}{a}. 

Consider now the whole of \figref{multifig}{}. It shows a transition from this striped Janus particle to a particle with alternating stripes. 
To understand this from the spectra, consider first \figref{multifig}{a}. 
The blue domain, giving the particle its Janus base pattern, comes from the minimum being dominated by a blue mode. 
When $\epsilon$ is increased, the amplitude of the blue mode decreases (\figref{multifig}{b}), which corresponds to a relative weakening of the phase separation of the blue particles.
Ultimately, the mode flips (\figref{multifig}{c}) to all states being in phase, causing this branch to become unphysical.
The global minimum among physical branches thus becomes $l=5$ and the lowest point involving the blue particles are now instead at $l=9$, in combination describing a base pattern that is red-yellow stripes decorated with blue stripes of double frequency.

Similar transitions can be found when the relative length parameters ($L_\alpha$) are varied.
\figref{lengthfig}{a-b} shows the transition from a Cerberus to a striped Janus particle.
In the first spectrum (\figref{lengthfig}{a}) the minimum is close to degenerate, with all branches minimized at $l=0$ due to the small differences in length parameters.
This causes all particle types to phase separate into distinct regions.
As the length differences between the shortest and the longer alkanethiols are increased (\figref{lengthfig}{b}), this degeneracy is lifted and the minimum of the red-yellow branch moves inwards.
This causes those particles to partially mix, again creating a striped Janus particle.

Increasing the lengths further causes a transition to a spotted Janus particle (\figref{lengthfig}{b-c}).
As discussed above, the striped Janus comes from the global minimum being dominated by a large blue region.
Here, as the lengths of the longer alkanethiols are increased, the relative energies of the two minima change until the red-yellow branch falls below the one giving the blue domain.
Additionally, the amplitude of the red mode increases.
This results in a pattern that is dominated by red spots (stripes if more red particles were available).
The former global minimum then describe the (continued) phase separation of the blue from the yellow particles, giving a spotted Janus particle.
Note the difference between the striped Janus, where the base pattern was Janus and the stripes a decoration, and this spotted Janus particle where the spots are primary and the Janus domains secondary.


In summary, we have presented a method for predicting the morphologies that self-assemble in general multi-component systems. We applied this method to a model of alkanethiol-on-gold systems and successfully predict its states and transitions.
We see three types of changes in morphologies: changes in the characteristic wavelength of the patterns due to smooth changes in the wavelength of the global minimum; abrupt changes in the morphologies due to destabilization of the global minimum (modes becoming unphysical); and abrupt changes in morphologies due to change of which local minimum is the global minimum.

The analytic method presented here can be used to design interactions that lead to self-assembly of specified geometries of patches and stripes. 
These patterns can be designed to promote the formation of target structures, \emph{e.g.}, four patches with tetrahedral configuration to get a diamond lattice, or a stripe at an angle from the pole to assemble polyhedra with a given number of faces. 
Preliminary results show that both these and other functional patterns are obtainable in the alkanethiol model and may therefore be realizable in experimental systems.

Many studies of patchy colloids concerns anisotropic particles rather than spherical.
For example, in practice it may be easier to design a desired surface morphology by altering the {\em shape} of the colloidal particle than by fine tuning the interactions between the components in the surface coating. 
The particle shape has indeed been the focus of many both simulation based~\cite{glotzer_anisotropy_2007,agarwal_mesophase_2011,damasceno_predictive_2012} and experimental~\cite{matijevic_monodispersed_1981,henzie_self-assembly_2012,rossi_cubic_2011} studies of entropy driven self-assembly, but adding surface coating to mitigate interactions between the particles gives new possibilities to control the assembly process.
It has been shown that such interactions can be much stronger between anisotropic particles than spherical ones~\cite{jones_nanoparticle_2011,glotzer_nanotechnology:_2012}.
The theory we present is not limited to spheres but can also be applied to more complicated geometries.
The more general form would be expressed in terms of eigenfunctions of the Laplace-Beltrami operator on curved surfaces, or in the case of polyhedra we would use the generalized operators for discrete differential geometry. 
This is an interesting direction for further investigations.

\noindent {\bf Acknowledgment.} OL and MNJ acknowledge support from the SuMo Biomaterials center of excellence.
We are also grateful to R. Bordes for discussions and comments.
%

\begin{thebibliography}{28}%
\makeatletter
\providecommand \@ifxundefined [1]{%
 \@ifx{#1\undefined}
}%
\providecommand \@ifnum [1]{%
 \ifnum #1\expandafter \@firstoftwo
 \else \expandafter \@secondoftwo
 \fi
}%
\providecommand \@ifx [1]{%
 \ifx #1\expandafter \@firstoftwo
 \else \expandafter \@secondoftwo
 \fi
}%
\providecommand \natexlab [1]{#1}%
\providecommand \enquote  [1]{``#1''}%
\providecommand \bibnamefont  [1]{#1}%
\providecommand \bibfnamefont [1]{#1}%
\providecommand \citenamefont [1]{#1}%
\providecommand \href@noop [0]{\@secondoftwo}%
\providecommand \href [0]{\begingroup \@sanitize@url \@href}%
\providecommand \@href[1]{\@@startlink{#1}\@@href}%
\providecommand \@@href[1]{\endgroup#1\@@endlink}%
\providecommand \@sanitize@url [0]{\catcode `\\12\catcode `\$12\catcode
  `\&12\catcode `\#12\catcode `\^12\catcode `\_12\catcode `\%12\relax}%
\providecommand \@@startlink[1]{}%
\providecommand \@@endlink[0]{}%
\providecommand \url  [0]{\begingroup\@sanitize@url \@url }%
\providecommand \@url [1]{\endgroup\@href {#1}{\urlprefix }}%
\providecommand \urlprefix  [0]{URL }%
\providecommand \Eprint [0]{\href }%
\providecommand \doibase [0]{http://dx.doi.org/}%
\providecommand \selectlanguage [0]{\@gobble}%
\providecommand \bibinfo  [0]{\@secondoftwo}%
\providecommand \bibfield  [0]{\@secondoftwo}%
\providecommand \translation [1]{[#1]}%
\providecommand \BibitemOpen [0]{}%
\providecommand \bibitemStop [0]{}%
\providecommand \bibitemNoStop [0]{.\EOS\space}%
\providecommand \EOS [0]{\spacefactor3000\relax}%
\providecommand \BibitemShut  [1]{\csname bibitem#1\endcsname}%
\let\auto@bib@innerbib\@empty
\bibitem [{\citenamefont {Jackson}\ \emph {et~al.}(2004)\citenamefont
  {Jackson}, \citenamefont {Myerson},\ and\ \citenamefont
  {Stellacci}}]{jackson_spontaneous_2004}%
  \BibitemOpen
  \bibfield  {author} {\bibinfo {author} {\bibfnamefont {A.~M.}\ \bibnamefont
  {Jackson}}, \bibinfo {author} {\bibfnamefont {J.~W.}\ \bibnamefont
  {Myerson}}, \ and\ \bibinfo {author} {\bibfnamefont {F.}~\bibnamefont
  {Stellacci}},\ }\href {\doibase 10.1038/nmat1116} {\bibfield  {journal}
  {\bibinfo  {journal} {Nat. Mater.}\ }\textbf {\bibinfo {volume} {3}},\
  \bibinfo {pages} {330} (\bibinfo {year} {2004})}\BibitemShut {NoStop}%
\bibitem [{\citenamefont {Bianchi}\ \emph {et~al.}(2011)\citenamefont
  {Bianchi}, \citenamefont {Blaak},\ and\ \citenamefont
  {Likos}}]{bianchi_patchy_2011}%
  \BibitemOpen
  \bibfield  {author} {\bibinfo {author} {\bibfnamefont {E.}~\bibnamefont
  {Bianchi}}, \bibinfo {author} {\bibfnamefont {R.}~\bibnamefont {Blaak}}, \
  and\ \bibinfo {author} {\bibfnamefont {C.~N.}\ \bibnamefont {Likos}},\
  }\href@noop {} {\bibfield  {journal} {\bibinfo  {journal} {Phys. Chem. Chem.
  Phys.}\ }\textbf {\bibinfo {volume} {13}},\ \bibinfo {pages} {6397} (\bibinfo
  {year} {2011})}\BibitemShut {NoStop}%
\bibitem [{\citenamefont {Yi}\ \emph {et~al.}(2013)\citenamefont {Yi},
  \citenamefont {Pine},\ and\ \citenamefont {Sacanna}}]{yi_progress_2013}%
  \BibitemOpen
  \bibfield  {author} {\bibinfo {author} {\bibfnamefont {G.-R.}\ \bibnamefont
  {Yi}}, \bibinfo {author} {\bibfnamefont {D.~J.}\ \bibnamefont {Pine}}, \ and\
  \bibinfo {author} {\bibfnamefont {S.}~\bibnamefont {Sacanna}},\ }\href@noop
  {} {\bibfield  {journal} {\bibinfo  {journal} {J.~Phys.:Condens. Matter}\
  }\textbf {\bibinfo {volume} {25}},\ \bibinfo {pages} {193101} (\bibinfo
  {year} {2013})}\BibitemShut {NoStop}%
\bibitem [{\citenamefont {Glotzer}\ \emph {et~al.}(2004)\citenamefont
  {Glotzer}, \citenamefont {Solomon},\ and\ \citenamefont
  {Kotov}}]{Glotzer:2004ce}%
  \BibitemOpen
  \bibfield  {author} {\bibinfo {author} {\bibfnamefont {S.~C.}\ \bibnamefont
  {Glotzer}}, \bibinfo {author} {\bibfnamefont {M.~J.}\ \bibnamefont
  {Solomon}}, \ and\ \bibinfo {author} {\bibfnamefont {N.~A.}\ \bibnamefont
  {Kotov}},\ }\href@noop {} {\bibfield  {journal} {\bibinfo  {journal} {AIChE
  J.}\ }\textbf {\bibinfo {volume} {50}},\ \bibinfo {pages} {2978} (\bibinfo
  {year} {2004})}\BibitemShut {NoStop}%
\bibitem [{\citenamefont {Pawar}\ and\ \citenamefont
  {Kretzschmar}(2010)}]{pawar_fabrication_2010}%
  \BibitemOpen
  \bibfield  {author} {\bibinfo {author} {\bibfnamefont {A.~B.}\ \bibnamefont
  {Pawar}}\ and\ \bibinfo {author} {\bibfnamefont {I.}~\bibnamefont
  {Kretzschmar}},\ }\href@noop {} {\bibfield  {journal} {\bibinfo  {journal}
  {Macromol. Rapid Commun.}\ }\textbf {\bibinfo {volume} {31}},\ \bibinfo
  {pages} {150} (\bibinfo {year} {2010})}\BibitemShut {NoStop}%
\bibitem [{\citenamefont {Doppelbauer}\ \emph {et~al.}(2012)\citenamefont
  {Doppelbauer}, \citenamefont {Noya}, \citenamefont {Bianchi},\ and\
  \citenamefont {Kahl}}]{Doppelbauer:2012kf}%
  \BibitemOpen
  \bibfield  {author} {\bibinfo {author} {\bibfnamefont {G.}~\bibnamefont
  {Doppelbauer}}, \bibinfo {author} {\bibfnamefont {E.~G.}\ \bibnamefont
  {Noya}}, \bibinfo {author} {\bibfnamefont {E.}~\bibnamefont {Bianchi}}, \
  and\ \bibinfo {author} {\bibfnamefont {G.}~\bibnamefont {Kahl}},\ }\href@noop
  {} {\bibfield  {journal} {\bibinfo  {journal} {Soft Matter}\ }\textbf
  {\bibinfo {volume} {8}},\ \bibinfo {pages} {7768} (\bibinfo {year}
  {2012})}\BibitemShut {NoStop}%
\bibitem [{\citenamefont {Singh}\ \emph {et~al.}(2007)\citenamefont {Singh},
  \citenamefont {Ghorai}, \citenamefont {Horsch}, \citenamefont {Jackson},
  \citenamefont {Larson}, \citenamefont {Stellacci},\ and\ \citenamefont
  {Glotzer}}]{singh_entropy-mediated_2007}%
  \BibitemOpen
  \bibfield  {author} {\bibinfo {author} {\bibfnamefont {C.}~\bibnamefont
  {Singh}}, \bibinfo {author} {\bibfnamefont {P.~K.}\ \bibnamefont {Ghorai}},
  \bibinfo {author} {\bibfnamefont {M.~A.}\ \bibnamefont {Horsch}}, \bibinfo
  {author} {\bibfnamefont {A.~M.}\ \bibnamefont {Jackson}}, \bibinfo {author}
  {\bibfnamefont {R.~G.}\ \bibnamefont {Larson}}, \bibinfo {author}
  {\bibfnamefont {F.}~\bibnamefont {Stellacci}}, \ and\ \bibinfo {author}
  {\bibfnamefont {S.~C.}\ \bibnamefont {Glotzer}},\ }\href {\doibase
  10.1103/PhysRevLett.99.226106} {\bibfield  {journal} {\bibinfo  {journal}
  {Phys. Rev. Lett.}\ }\textbf {\bibinfo {volume} {99}},\ \bibinfo {pages}
  {226106} (\bibinfo {year} {2007})}\BibitemShut {NoStop}%
\bibitem [{\citenamefont {{Pons-Siepermann}}\ and\ \citenamefont
  {Glotzer}(2012)}]{pons-siepermann_design_2012a}%
  \BibitemOpen
  \bibfield  {author} {\bibinfo {author} {\bibfnamefont {I.~C.}\ \bibnamefont
  {{Pons-Siepermann}}}\ and\ \bibinfo {author} {\bibfnamefont {S.~C.}\
  \bibnamefont {Glotzer}},\ }\href {\doibase 10.1039/C2SM00014H} {\bibfield
  {journal} {\bibinfo  {journal} {Soft Matter}\ }\textbf {\bibinfo {volume}
  {8}},\ \bibinfo {pages} {6226} (\bibinfo {year} {2012})}\BibitemShut
  {NoStop}%
\bibitem [{\citenamefont {Pons-Siepermann}\ and\ \citenamefont
  {Glotzer}(2012)}]{pons-siepermann_design_2012b}%
  \BibitemOpen
  \bibfield  {author} {\bibinfo {author} {\bibfnamefont {I.~C.}\ \bibnamefont
  {Pons-Siepermann}}\ and\ \bibinfo {author} {\bibfnamefont {S.~C.}\
  \bibnamefont {Glotzer}},\ }\href {\doibase 10.1021/nn300059x} {\bibfield
  {journal} {\bibinfo  {journal} {ACS Nano}\ }\textbf {\bibinfo {volume} {6}},\
  \bibinfo {pages} {3919} (\bibinfo {year} {2012})}\BibitemShut {NoStop}%
\bibitem [{\citenamefont {Cesbron}\ \emph {et~al.}(2012)\citenamefont
  {Cesbron}, \citenamefont {Shaw}, \citenamefont {Birchall}, \citenamefont
  {Free},\ and\ \citenamefont {L\'evy}}]{cesbron_stripy_2012}%
  \BibitemOpen
  \bibfield  {author} {\bibinfo {author} {\bibfnamefont {Y.}~\bibnamefont
  {Cesbron}}, \bibinfo {author} {\bibfnamefont {C.~P.}\ \bibnamefont {Shaw}},
  \bibinfo {author} {\bibfnamefont {J.~P.}\ \bibnamefont {Birchall}}, \bibinfo
  {author} {\bibfnamefont {P.}~\bibnamefont {Free}}, \ and\ \bibinfo {author}
  {\bibfnamefont {R.}~\bibnamefont {L\'evy}},\ }\href {\doibase
  10.1002/smll.201001465} {\bibfield  {journal} {\bibinfo  {journal} {Small}\
  }\textbf {\bibinfo {volume} {8}},\ \bibinfo {pages} {3714} (\bibinfo {year}
  {2012})}\BibitemShut {NoStop}%
\bibitem [{\citenamefont {Yu}\ and\ \citenamefont
  {Stellacci}(2012)}]{yu_response_2012}%
  \BibitemOpen
  \bibfield  {author} {\bibinfo {author} {\bibfnamefont {M.}~\bibnamefont
  {Yu}}\ and\ \bibinfo {author} {\bibfnamefont {F.}~\bibnamefont {Stellacci}},\
  }\href {\doibase 10.1002/smll.201202322} {\bibfield  {journal} {\bibinfo
  {journal} {Small}\ }\textbf {\bibinfo {volume} {8}},\ \bibinfo {pages} {3720}
  (\bibinfo {year} {2012})}\BibitemShut {NoStop}%
\bibitem [{\citenamefont {Likos}(2001)}]{likos_effective_2001}%
  \BibitemOpen
  \bibfield  {author} {\bibinfo {author} {\bibfnamefont {C.~N.}\ \bibnamefont
  {Likos}},\ }\href@noop {} {\bibfield  {journal} {\bibinfo  {journal} {Phys.
  Rep.}\ }\textbf {\bibinfo {volume} {348}},\ \bibinfo {pages} {267} (\bibinfo
  {year} {2001})}\BibitemShut {NoStop}%
\bibitem [{Note1()}]{Note1}%
  \BibitemOpen
  \bibinfo {note} {E. Edlund, O. Lindgren, and M. Nilsson Jacobi, in
  preparation (2013)}\BibitemShut {NoStop}%
\bibitem [{\citenamefont {Baxter}(1982)}]{baxter_exactly_1982}%
  \BibitemOpen
  \bibfield  {author} {\bibinfo {author} {\bibfnamefont {R.~J.}\ \bibnamefont
  {Baxter}},\ }\href@noop {} {\emph {\bibinfo {title} {{Exactly Solved Models
  in Statistical Mechanics}}}}\ (\bibinfo  {publisher} {Academic Press},\
  \bibinfo {address} {London},\ \bibinfo {year} {1982})\ p.\ \bibinfo {pages}
  {322}\BibitemShut {NoStop}%
\bibitem [{Note2()}]{Note2}%
  \BibitemOpen
  \bibinfo {note} {Here the careful reader notes that it is impossible to
  construct a uniform mesh on a sphere (except for the meshes of 4, 6, 8, 12,
  or 20 points corresponding to the Platonic solids). However, our theory does
  not need a uniform mesh. For the numerical spectra calculations, we use a
  configuration obtained by Monte Carlo annealing~\cite {Note4} for $N=880$
  particles on a sphere with radius $9$. Using different configurations makes
  negligible difference for the spectra.}\BibitemShut {Stop}%
\bibitem [{\citenamefont {Edlund}\ and\ \citenamefont {{Nilsson
  Jacobi}}(2010)}]{edlund_universality_2010}%
  \BibitemOpen
  \bibfield  {author} {\bibinfo {author} {\bibfnamefont {E.}~\bibnamefont
  {Edlund}}\ and\ \bibinfo {author} {\bibfnamefont {M.}~\bibnamefont {{Nilsson
  Jacobi}}},\ }\href@noop {} {\bibfield  {journal} {\bibinfo  {journal} {Phys.
  Rev. Lett.}\ }\textbf {\bibinfo {volume} {105}},\ \bibinfo {pages} {137203}
  (\bibinfo {year} {2010})}\BibitemShut {NoStop}%
\bibitem [{\citenamefont {Nussinov}(2001)}]{nussinov_commensurate_2001}%
  \BibitemOpen
  \bibfield  {author} {\bibinfo {author} {\bibfnamefont {Z.}~\bibnamefont
  {Nussinov}},\ }\href@noop {} {\bibfield  {journal} {\bibinfo  {journal}
  {arxiv:cond-mat/0105253}\ } (\bibinfo {year} {2001})}\BibitemShut {NoStop}%
\bibitem [{Note4()}]{Note4}%
  \BibitemOpen
  \bibinfo {note} {All particle configurations displayed in this Letter are
  generated using Monte Carlo annealing of $N=1500$ particles restricted to the
  surface of a sphere with radios $R=12\sigma _0$. Trial moves consists of
  either, with equal probability, interchanging the position of two particles
  or choosing a new random position on the surface of the sphere for a single
  particle. The annealing is done for $10^4$ sweeps at each of the temperatures
  $T\in \protect \{1000, 100, 10, 3, 1, 0.3, 0.1\protect \}$, where a sweep is
  $N$ trials.}\BibitemShut {Stop}%
\bibitem [{Note3()}]{Note3}%
  \BibitemOpen
  \bibinfo {note} {The reader may note the similarity between our
  diagonalization procedure and the standard approach to calculating lattice
  phonon spectra~\cite {ashcroft_solid_1976}, though the interpretation of the
  results is different.}\BibitemShut {Stop}%
\bibitem [{\citenamefont {Glotzer}\ and\ \citenamefont
  {Solomon}(2007)}]{glotzer_anisotropy_2007}%
  \BibitemOpen
  \bibfield  {author} {\bibinfo {author} {\bibfnamefont {S.~C.}\ \bibnamefont
  {Glotzer}}\ and\ \bibinfo {author} {\bibfnamefont {M.~J.}\ \bibnamefont
  {Solomon}},\ }\href@noop {} {\bibfield  {journal} {\bibinfo  {journal} {Nat.
  Mater.}\ }\textbf {\bibinfo {volume} {6}},\ \bibinfo {pages} {557} (\bibinfo
  {year} {2007})}\BibitemShut {NoStop}%
\bibitem [{\citenamefont {Agarwal}\ and\ \citenamefont
  {Escobedo}(2011)}]{agarwal_mesophase_2011}%
  \BibitemOpen
  \bibfield  {author} {\bibinfo {author} {\bibfnamefont {U.}~\bibnamefont
  {Agarwal}}\ and\ \bibinfo {author} {\bibfnamefont {F.~A.}\ \bibnamefont
  {Escobedo}},\ }\href@noop {} {\bibfield  {journal} {\bibinfo  {journal} {Nat.
  Mater.}\ }\textbf {\bibinfo {volume} {10}},\ \bibinfo {pages} {230} (\bibinfo
  {year} {2011})}\BibitemShut {NoStop}%
\bibitem [{\citenamefont {Damasceno}\ \emph {et~al.}(2012)\citenamefont
  {Damasceno}, \citenamefont {Engel},\ and\ \citenamefont
  {Glotzer}}]{damasceno_predictive_2012}%
  \BibitemOpen
  \bibfield  {author} {\bibinfo {author} {\bibfnamefont {P.~F.}\ \bibnamefont
  {Damasceno}}, \bibinfo {author} {\bibfnamefont {M.}~\bibnamefont {Engel}}, \
  and\ \bibinfo {author} {\bibfnamefont {S.~C.}\ \bibnamefont {Glotzer}},\
  }\href@noop {} {\bibfield  {journal} {\bibinfo  {journal} {Science}\ }\textbf
  {\bibinfo {volume} {337}},\ \bibinfo {pages} {453} (\bibinfo {year}
  {2012})}\BibitemShut {NoStop}%
\bibitem [{\citenamefont {Matijevic}(1981)}]{matijevic_monodispersed_1981}%
  \BibitemOpen
  \bibfield  {author} {\bibinfo {author} {\bibfnamefont {E.}~\bibnamefont
  {Matijevic}},\ }\href {\doibase 10.1021/ar00061a004} {\bibfield  {journal}
  {\bibinfo  {journal} {Acc. Chem. Res.}\ }\textbf {\bibinfo {volume} {14}},\
  \bibinfo {pages} {22} (\bibinfo {year} {1981})}\BibitemShut {NoStop}%
\bibitem [{\citenamefont {Henzie}\ \emph {et~al.}(2012)\citenamefont {Henzie},
  \citenamefont {Gr\"unwald}, \citenamefont {Widmer-Cooper}, \citenamefont
  {Geissler},\ and\ \citenamefont {Yang}}]{henzie_self-assembly_2012}%
  \BibitemOpen
  \bibfield  {author} {\bibinfo {author} {\bibfnamefont {J.}~\bibnamefont
  {Henzie}}, \bibinfo {author} {\bibfnamefont {M.}~\bibnamefont {Gr\"unwald}},
  \bibinfo {author} {\bibfnamefont {A.}~\bibnamefont {Widmer-Cooper}}, \bibinfo
  {author} {\bibfnamefont {P.~L.}\ \bibnamefont {Geissler}}, \ and\ \bibinfo
  {author} {\bibfnamefont {P.}~\bibnamefont {Yang}},\ }\href {\doibase
  10.1038/nmat3178} {\bibfield  {journal} {\bibinfo  {journal} {Nat. Mater.}\
  }\textbf {\bibinfo {volume} {11}},\ \bibinfo {pages} {131} (\bibinfo {year}
  {2012})}\BibitemShut {NoStop}%
\bibitem [{\citenamefont {Rossi}\ \emph {et~al.}(2011)\citenamefont {Rossi},
  \citenamefont {Sacanna}, \citenamefont {Irvine}, \citenamefont {Chaikin},
  \citenamefont {Pine},\ and\ \citenamefont {Philipse}}]{rossi_cubic_2011}%
  \BibitemOpen
  \bibfield  {author} {\bibinfo {author} {\bibfnamefont {L.}~\bibnamefont
  {Rossi}}, \bibinfo {author} {\bibfnamefont {S.}~\bibnamefont {Sacanna}},
  \bibinfo {author} {\bibfnamefont {W.~T.~M.}\ \bibnamefont {Irvine}}, \bibinfo
  {author} {\bibfnamefont {P.~M.}\ \bibnamefont {Chaikin}}, \bibinfo {author}
  {\bibfnamefont {D.~J.}\ \bibnamefont {Pine}}, \ and\ \bibinfo {author}
  {\bibfnamefont {A.~P.}\ \bibnamefont {Philipse}},\ }\href {\doibase
  10.1039/C0SM01246G} {\bibfield  {journal} {\bibinfo  {journal} {Soft Matter}\
  }\textbf {\bibinfo {volume} {7}},\ \bibinfo {pages} {4139} (\bibinfo {year}
  {2011})}\BibitemShut {NoStop}%
\bibitem [{\citenamefont {Jones}\ \emph {et~al.}(2011)\citenamefont {Jones},
  \citenamefont {Macfarlane}, \citenamefont {Prigodich}, \citenamefont
  {Patel},\ and\ \citenamefont {Mirkin}}]{jones_nanoparticle_2011}%
  \BibitemOpen
  \bibfield  {author} {\bibinfo {author} {\bibfnamefont {M.~R.}\ \bibnamefont
  {Jones}}, \bibinfo {author} {\bibfnamefont {R.~J.}\ \bibnamefont
  {Macfarlane}}, \bibinfo {author} {\bibfnamefont {A.~E.}\ \bibnamefont
  {Prigodich}}, \bibinfo {author} {\bibfnamefont {P.~C.}\ \bibnamefont
  {Patel}}, \ and\ \bibinfo {author} {\bibfnamefont {C.~A.}\ \bibnamefont
  {Mirkin}},\ }\href {\doibase 10.1021/ja206777k} {\bibfield  {journal}
  {\bibinfo  {journal} {J.~Amer. Chem. Soc.}\ }\textbf {\bibinfo {volume}
  {133}},\ \bibinfo {pages} {18865} (\bibinfo {year} {2011})}\BibitemShut
  {NoStop}%
\bibitem [{\citenamefont {Glotzer}(2012)}]{glotzer_nanotechnology:_2012}%
  \BibitemOpen
  \bibfield  {author} {\bibinfo {author} {\bibfnamefont {S.~C.}\ \bibnamefont
  {Glotzer}},\ }\href {\doibase 10.1038/481450a} {\bibfield  {journal}
  {\bibinfo  {journal} {Nature}\ }\textbf {\bibinfo {volume} {481}},\ \bibinfo
  {pages} {450} (\bibinfo {year} {2012})}\BibitemShut {NoStop}%
\bibitem [{\citenamefont {Ashcroft}\ and\ \citenamefont
  {Mermin}(1976)}]{ashcroft_solid_1976}%
  \BibitemOpen
  \bibfield  {author} {\bibinfo {author} {\bibfnamefont {N.~W.}\ \bibnamefont
  {Ashcroft}}\ and\ \bibinfo {author} {\bibfnamefont {N.~D.}\ \bibnamefont
  {Mermin}},\ }\enquote {\bibinfo {title} {{Solid State Physics}},}\ \
  (\bibinfo  {publisher} {Brooks/Cole},\ \bibinfo {address} {Belmont},\
  \bibinfo {year} {1976})\ p.\ \bibinfo {pages} {442}\BibitemShut {NoStop}%
\end{thebibliography}


%


\end{document}